\documentclass[reprint,amsmath,amssymb,aps]{revtex4-2}
\usepackage{soul}%st
\usepackage{graphicx}
\usepackage{dcolumn}
\usepackage{bm}
\usepackage{appendix}
\usepackage{float}
\usepackage{xcolor,cancel}
\usepackage{xcolor}

\begin{document}

\preprint{APS/123-QED}

\title{The equilibrium configurations of neutron stars in the optimized $f(R,T)$ gravity}

\author{J.T. Quartuccio}
 \email{jon.quartuccio@gmail.com}
\affiliation{Laborat\'orio de F\'isica Te\'orica e Computacional (LFTC),
 Universidade Cidade de S\~ao Paulo (UNICID) - Rua Galv\~ao Bueno 868, 01506-000 S\~ao Paulo, Brazil}
 \author{P.H.R.S. Moraes}
 \email{moraes.phrs@gmail.com}
\affiliation{Laborat\'orio de F\'isica Te\'orica e Computacional (LFTC),
 Universidade Cidade de S\~ao Paulo (UNICID) - Rua Galv\~ao Bueno 868, 01506-000 S\~ao Paulo, Brazil}
 \author{G.N. Zeminiani}
 \email{guilherme.zeminiani@gmail.com}
\affiliation{Laborat\'orio de F\'isica Te\'orica e Computacional (LFTC),
 Universidade Cidade de S\~ao Paulo (UNICID) - Rua Galv\~ao Bueno 868, 01506-000 S\~ao Paulo, Brazil} 
 \author{M.M. Lapola}
 \email{marcelo.lapola@gmail.com}
 \affiliation{Núcleo Comum de Engenharia, Centro Universitário Fundação Hermínio Ometto, Av. Dr. Maximiliano Baruto, 500 - Jardim Universitario, Araras - SP, 13607-339, Brazil}

\begin{abstract}
\begin{center}
\end{center}

  We construct equilibrium configurations for neutron stars using a specific $f(R,T)$ 
functional form, recently derived through gaussian process applied to measurements of the Hubble parameter. By construction, this functional form serves as an alternative explanation for cosmic acceleration, circumventing the cosmological constant problem. Here, we aim to examine its applicability within the stellar regime. In doing so, we seek to contribute to the modified gravity literature by applying the same functional form of a given gravity theory across highly distinct regimes. Our results demonstrate that equilibrium configurations of neutron stars can be obtained within this theory, with the energy density and maximum mass slightly exceeding those predicted by General Relativity. Additionally, we show that the value of some parameters in the $f(R,T)$ functional form must differ from those obtained in cosmological configurations, suggesting a potential scale-dependence for these parameters. We propose that further studies apply this functional form across different regimes to more thoroughly assess this possible dependence.\\


\end{abstract}  

\maketitle

\section{Introduction}\label{sec:int}

Modified gravity is a highly active area of research in physics, as it has the potential to explain the cosmic acceleration without relying on the cosmological constant. The advantage of that is the evasion of the infamous cosmological constant problem, which is the tremendous discrepancy between the theoretically predicted and observed values of the vacuum energy density \cite{weinberg/1989,adler/1995,peebles/2003,padmanabhan/2003}.

The operational framework of modified gravity is outlined in the following. General Relativity field equations in the presence of the cosmological constant can be obtained from the variation of the following action 

\begin{equation}\label{i1}
    S=\frac{1}{16\pi}\int d^4x\sqrt{-g}(R-2\Lambda)+\int d^4x\sqrt{-g}\mathcal{L}_m,
\end{equation}
in which $g$ is the metric determinant, $R$ is the Ricci or curvature scalar, $\Lambda$ is the cosmological constant, $\mathcal{L}_m$ is the matter lagrangian density and natural units are assumed. Modified gravity (usually) neglects $\Lambda$ and substitutes $R$ by a function of it, namely, $f(R)$, and even of other scalars. The variation of these augmented actions yields field equations with extra terms in comparison with General Relativity field equations, and those extra terms may be capable of explaining the cosmic acceleration in accordance with observations. Some successful examples of that are presented, e.g., in \cite{ishak/2006,easson/2004,lue/2004,solanki/2021}. 

The extra terms of the modified gravity field equations must be negligible in regimes where General Relativity is already probed. In other words, for a modified gravity theory to be acceptable, it must converge to the results of General Relativity in the regimes where the latter has already been experimentally validated. Some severe solar system constraints were put to modified gravity in \cite{chakraborty/2014,de_felice/2009,chan/2023}. 

It is an unpleasant current feature of modified gravity theory applications that usually there is no concern with the behaviour of a particular model in different regimes of application. That is, when one applies a particular modified gravity to a particular field of applications (equilibrium configurations of compact objects, solar system, galactic dynamics, structure formation, cosmological models etc.), there is usually no consideration of the outcomes obtained in the remaining applications. Let us exemplify such an argumentation in the following.

In a paper by Capozziello and collaborators \cite{capozziello/2007}, the rotation curves of $15$ low surface brightness galaxies were examined, showing no need for dark matter, in the $f(R)=f_0R^n$ gravity, with $f_0$ a constant that accounts for the correct units and $n=3.5$. Later, Faraoni showed that in order for the scalar field in the scalar-tensor version of $R^n$ gravity to be non-tachyonic, one must have $1\leq n\leq2$ \cite{faraoni/2011}. More dramatically, by writing  $n=1+\delta$, the precession of the perihelion of Mercury yields $\delta=(2.7\pm4.5)\times10^{-19}$ \cite{clifton/2005,clifton/2006,barrow/2006}, a stringent limit often ignored in $R^n$ gravity (check References \cite{leon/2011,bisabr/2010,capozziello/2008,corda/2008,corda/2008b,corda/2007,martins/2007,mendoza/2007}, among others). 

Allemandi and Ruggiero have written a quite interesting paper in which they use gravitational redshift, light deflection, gravitational time-delay and geodesic precession to constrain different $f(R)$ gravity models \cite{allemandi/2007}. Remarkably, they obtained that the estimated values of the $f(R)$ functional form parameters are orders of magnitude bigger than the values obtained in the cosmological framework.

As one would expect, this is not a feature of $f(R)$ gravity theory only. Exactly motivated by severe weak field constraints on $f(R)$ theory that, as a matter of fact, rule out most of the suggested models \cite{chiba/2003,chiba/2007,erickcek/2006,nojiri/2008,capozziello/2007b,capozziello/2008b,olmo/2007}, Harko and collaborators proposed the $f(R,T)$ gravity theory \cite{harko/2011}, for which $T$ is the trace of the energy-momentum tensor. The motivations to insert terms on $T$ in the gravitational action will be provided next. For now, let us mention that the modified Tolman-Oppenheimer-Volkoff (TOV) equation was derived for the first time for the $f(R,T)$ gravity in \cite{moraes/2016}. The functional form chosen for $f(R,T)$ was $f(R,T)=R+2\lambda T$, with constant $\lambda$, and such a constant presented different values - around one order of magnitude - when solutions were obtained for neutron stars and strange stars, being lower in the former case. Later, the same approach was applied to white dwarfs and a lower limit for $\lambda$ was found, namely $\vert\lambda\vert>3\times10^{-4}$ \cite{carvalho/2017}.

The examples on the above regard are numerous and spread to several modified gravity theories. They point to an incoherence in the results of a given theory when applied to different regimes or to a hidden scale-dependence of these ``constants''.

It is the purpose of the present article to verify the neutron stars equilibrium configurations viability of a particular functional form of the $f(R,T)$ gravity which is well succeeded, by construction, in the cosmological scenario. We will work with the functional form found by Fortunato et al. \cite{fortunato/2024}. In \cite{fortunato/2024}, the $f(R,T)$ functional form was constructed through gaussian process \cite{seikel/2012}, that was applied through the use of a series of measurements of the Hubble parameter. The result obtained was

\begin{equation}\label{i2}
    f(T)=\alpha T^2+A\tanh[\lambda(T+T_0)]+\beta T+\gamma.
\end{equation}
More details on this result as well as the values obtained for the constants $\alpha, A, \lambda, T_0, \beta$ and $\gamma$ will be given in the following.

The article is organized as follows. In Section \ref{sec:frt} we present the basis of the $f(R,T)$ gravity. In Section \ref{sec:tov}, we derive the TOV-like equation. In Section \ref{sec:res} we present our results, which are obtained by particularizing $f(T)$ as according to Eq.\eqref{i2}. Those are discussed in Section \ref{sec:dcr}, in which we also provide some perspectives for applications of \eqref{i2}.

\section{$f(R,T)$ gravity}\label{sec:frt}

The $f(R,T)$ gravity is an extension of General Relativity Theory proposed by Harko et al. \cite{harko/2011}. The theory assumes that the gravitational action depends on a generic function of both $R$ and $T$, namely, the Ricci scalar and the trace of the energy-momentum tensor, respectively. Taking $\mathcal{L}_m$ as the matter lagrangian density, the total action of $f(R,T)$ gravity is written as

\begin{equation}\label{action}
	S=\int d^4x\sqrt{-g}\left[\frac{f(R,T)}{16\pi}+\mathcal{L}_m\right].
\end{equation}
%in which $g$ is the determinant of the metric tensor $g_{\mu \nu}$ and natural units are assumed. 

Applying the principle of least action to Equation (\ref{action}) we obtain the $f(R,T)$ field equations
\begin{equation}\label{fieldeq}
	(R_{\mu \nu}+g_{\mu \nu}\Box - \nabla_\mu \nabla_\nu)f_R-\frac{1}{2}fg_{\mu \nu}= 8\pi T_{\mu \nu}+f_T(T_{\mu \nu}+pg_{\mu \nu}),
\end{equation}
in which $R_{\mu \nu}$ is the Ricci tensor, $g_{\mu\nu}$ is the metric tensor,  $f_R=f_R(R,T)\equiv\partial f(R,T)/\partial R$, $f=f(R,T)$, $T_{\mu\nu}\equiv(-2/\sqrt{-g})\delta(\sqrt{-g}\mathcal{L}_m)/\delta g_{\mu\nu}$ is the energy-momentum tensor (defined as usually), $f_T=f_T(R,T)\equiv\partial f(R,T)/\partial T$ and we have taken $\mathcal{L}_m=-p$, with $p$ being the pressure. 

It is usual to express the $f(R,T)$ functional in the form $f(R,T) = R + f(T)$, that is, keeping the same $R$-dependence as in General Relativity Theory, but considering terms that depend on $T$. On this regard, the $T$-dependence motivation in the $f(R,T)$ gravity  will be briefly discussed in the following. 

The dependence on $T$ may be induced by quantum effects (actually, in a similar way as the dependence on $\Lambda$ in Equation \eqref{i1}), specifically, the trace anomaly \cite{birrell/1982}. In quantum field theories, the trace anomaly refers to the situation where the trace of the energy-momentum tensor, which in a classical field theory might be zero, becomes non-zero due to quantum corrections. In other words, the scale symmetry is broken at the quantum level, introducing a residual effect from vacuum energy that can interact with geometry. On this regard, it is worth mentioning that the explanation for the acceleration in the expansion of the universe has been proposed through trace anomaly in the General Relativity context by Sch\"{u}tzhold \cite{schutzhold/2002}.

Assuming $f(R,T)=R+f(T)$ in (\ref{fieldeq}) yields
\begin{equation}\label{GfRT}
	G_{\mu \nu} = 8\pi T_{\mu \nu} + f_T(T_{\mu \nu} + p g_{\mu \nu}) + \frac{1}{2}f(T)g_{\mu \nu},
\end{equation}
with $G_{\mu\nu}$ being the Einstein tensor.

By applying the covariant divergence to \eqref{GfRT}, we obtain
\begin{equation}\label{covariant}
        \nabla^\mu T_{\mu \nu} = \frac{-f_T}{8\pi+f_T}\left[(T_{\mu\nu}+pg_{\mu\nu})\nabla^\mu\ln f_T+g_{\mu\nu}\nabla^\mu\left(p+\frac{T}{2}\right)\right].
\end{equation}

\section{TOV-like equation in $f(R,T)$ gravity}\label{sec:tov}

To derive the hydrostatic equilibrium equation in the $f(R,T)$ gravity, we start by applying a spherically symmetric metric and the energy-momentum tensor of a perfect fluid to Equation (\ref{GfRT}). Those are given, respectively, by
\begin{equation}\label{metric}
	ds^2 = e^{\upsilon(r)}dt^2 - e^{\omega(r)}dr^2 - r^2(d\theta^2 + \sin^2\theta d\phi^2),
\end{equation}
with $\upsilon(r)$ and $\omega(r)$ being metric potentials, and

\begin{equation}\label{tov1}
    T_{\mu\nu}=\text{diag}(\rho,-p,-p,-p),
\end{equation}
with $\rho$ being the matter-energy density of the star.

%From (\ref{metric}), the non-null components of the Einstein tensor are given by

%\begin{equation}
%    G_{22} = \frac{1}{4}e^{-\omega}r[r\upsilon'^2 - 2\omega' + \upsilon'(2-r\omega')+2r\upsilon''],
%\end{equation}
%\begin{equation}
%    G_{33} = G_{22}\sin^2 \theta,
%\end{equation}
%from which primes represent derivatives with respect to $r$.

From \eqref{metric} and \eqref{tov1}, we have that

\begin{equation}\label{nablaT}
	\nabla^\mu T_{\mu \nu} = p' +  (\rho + p)\frac{\upsilon'}{2},
\end{equation}
with a prime indicating radial derivative. Replacing (\ref{nablaT}) in (\ref{covariant}) yields
\begin{equation}\label{plinha2}
\begin{split}
    	p' + (\rho + p)\upsilon' =\frac{f_T}{8\pi + f_T}(p'-\rho').
\end{split}
\end{equation}

Applying the $G_{11}$ component, namely,

\begin{equation}
    G_{11} = \frac{1 - e^\omega - r\upsilon'}{r^2},
\end{equation}
in (\ref{GfRT}), we find
\begin{equation}\label{upsilonlinha}
	\upsilon' = 8\pi pre^\omega  - \frac{1}{2}fre^\omega -\frac{1+e^\omega}{r},
\end{equation}
and inserting Equation (\ref{upsilonlinha}) in (\ref{plinha2}) finally yields the $f(R,T)$ gravity TOV-like equation
\begin{equation}\label{TOVfRT2}
p' = -(\rho + p) \frac{\left[4\pi p - \frac{f(T)}{4}\right]r + \frac{m}{r^2}}{\left(1 - \frac{2m}{r}\right)\left[1 - \frac{f_T}{2(8\pi + f_T)}\left(1 - \frac{d\rho}{dp}\right)\right]}.
\end{equation}
It is evident that when $f(T) = 0$, the ``standard'' TOV equation is recovered \cite{open}.

To obtain the mass equation, we apply the $G_{00}$ component, namely 

\begin{equation}
    G_{00} = \frac{e^{\upsilon - \omega}(-1 + e^\omega + r\omega')}{r^2},
\end{equation}
in Equation (\ref{GfRT}), to obtain

\begin{equation}\label{mass}
		m' = 4\pi r^2 \rho + \frac{r^2}{2}\left[f_T(\rho + p)+ \frac{f(T)}{2}\right].
\end{equation}
Analogously, by making $f(T)=0$, the standard mass equation is retrieved. 

In this work, we consider the $f(T)$ functional form obtained by Fortunato et al. in \cite{fortunato/2024} and given by Eq.\eqref{i2}, i.e.,

\begin{equation*}
    f(T)=\alpha T^2+A\tanh[\lambda(T+T_0)]+\beta T+\gamma.
\end{equation*}
In such a reference, the values obtained for the above constants are $\alpha=-1.83\times10^{-5}$, $A=-1.05\times10^4$, $\lambda=-2.39\times10^3$, $T_0=2.58\times10^3$, $\beta=-2.99$ and $\gamma=-1.61\times10^4$. Recall that while most $f(R,T)$ functions in the literature were proposed {\it ad hoc}, Eq.\eqref{i2} was constructed by using gaussian process and a series of measurements of the Hubble parameter. 

\section{Neutron star solutions}\label{sec:res}

The stellar structure configurations are analyzed using the polytropic equation of state, given by $p=K \rho^{\Gamma}$, where $K = 1.24\times 10^{-4}$ (fm$^3$/MeV) and $\Gamma = 2$ \cite{Horedt, Chima, TOV}. We solve the structure equations (\ref{TOVfRT2}) and (\ref{mass}) by numerically integrating from the center of the star ($r= 0$) to its surface ($r=R$) using the Euler algorithm adapted from \cite{TOV}. % K está sem unidades, Gamma parece grande

The solutions begin with the values at $r = 0$, namely $\rho_0 = \rho_c = 10^{14}$ g/cm³ and $m(0) = 0$, and ends when the star surface is found, that is, $p(R) = 0$ \cite{TOV}.

For the constants of the $f(R,T)$ functional, we notice that $A$, $\beta$ and $\gamma$ play a crucial role in our analysis. Unlike the values obtained by Fortunato et al., they  must be close to zero in order to accurately describe a neutron star, while $\alpha$, $\lambda$ and $T_0$ keep the same values as in the cosmological case \cite{fortunato/2024}. For $A = 5.0\times 10^{-4}$ and $\beta = \gamma = 1.05\times 10^{-3}$, we obtain a maximum mass of 1.67M$_{\odot}$, according to Fig.\ref{fig:enter-label} below.

%\begin{figure}[H]
%\centering
%\includegraphics[width=8.5cm]{-1e-3.png}
%\includegraphics[width=8.5cm]{0.png} 
%\caption{\label{fig1} In the top panel, we display the mass-radius relation for $\beta = %-1\times10^{-3}$, where the maximum mass reaches $1.56$M$_\odot$ with a radius of $10.80$ km. %In the bottom panel, assuming $\beta = 0$, the maximum mass obtained is $1.61$M$_\odot$ for a %radius of $11.10$ km.}
%\end{figure}

%\begin{figure}[H]
 %   \centering
  %  \includegraphics[scale=0.55]{1e-3.png}
   % \caption{The maximum mass is reached for $\beta = 1\times10^{-3}$ , yielding a mass of $1.67$M$_\odot$ at a radius of $11.56$ km. Beyond this point, a sharp decline in mass is observed as the radius increases further. This behavior is characteristic of the solutions to the TOV equations, marking the transition between stable and unstable configurations.}
  %  \label{fig:enter-label}
%\end{figure}

\begin{figure}[H]
    \centering
    \includegraphics[scale=0.55]{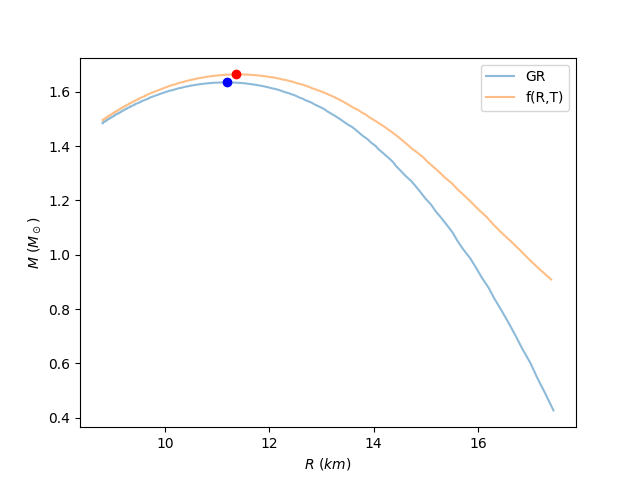}
    \caption{Comparison between neutron star mass predictions in the framework of General Relativity (blue line) and the optimized $f(R,T)$ gravity (orange line). The highlighted points indicate the maximum masses obtained for each model.}
    \label{fig:enter-label}
\end{figure}

\begin{figure}[H]
    \centering
    \includegraphics[scale=0.55]{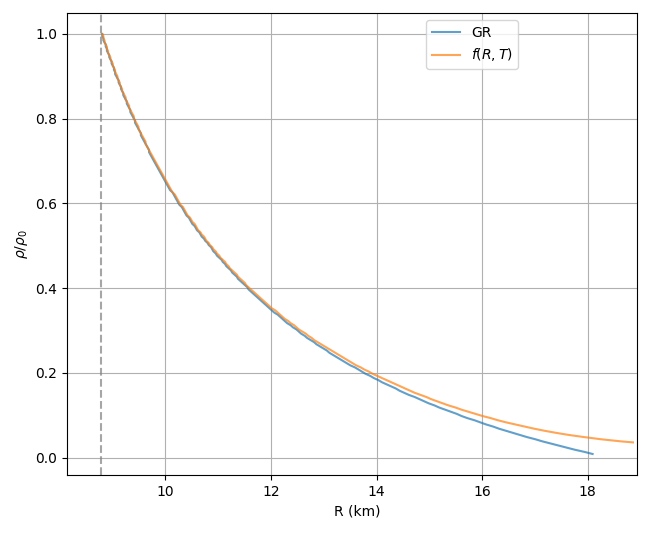}
    \caption{Energy density $\rho/\rho_0$ as a function of the radius for both General Relativity (blue line) and optimized $f(R,T)$ gravity (orange line).}
    \label{fig:enter-label2}
\end{figure}

In Fig.\ref{fig:enter-label2} below we show the energy density vs. radius for General Relativity and optimized $f(R,T)$ theory. The comparison reveals differences in the predictions of each model regarding the internal structure of neutron stars. Evidently, in the $f(R,T)$ theory, it is possible to achieve higher densities compared to General Relativity, which allows for an increase in the maximum mass of the star.

% esta seção está bem confusa. por que há duas figuras na figura 1? porque não colocar as duas curvas na mesma figura? e o que significa essa degenerescência no painel superior? ou seja, no intervalo 13-15 para o raio, há dois valores para massa. como escolher um em detrimento do outro? o que você quis dizer para o leitor quando colocou estes dois valores para beta? em seguida, voce plota uma terceira relacao massa-raio para um terceiro valor de beta, diferente dos dois anteriores. para raios proximos a 22, voce tem uma serie de valores para massa. uma baita degenerescencia. isto nao pode aparecer. e finalmente a figura 4, que poderia ser a unica figura do artigo (até agora), se fosse retirado esse mal comportamento dela a altos raios. uma outra coisa, geralmente também se plota a densidade central contra o raio e/ou contra a massa. você não construiu esses plots e tampouco informou que valor está sendo usado para a densidade central. consegue fazer isso?

\section{Discussion and concluding remarks}\label{sec:dcr}

We discussed in the Introduction section a setback that modified gravity theories face when applied to different gravitational regimes. Apparently, the theory parameters need to adapt their value according to the system the theory is applied. As those parameters are ultimately appearing in a theory's action, this scale problem represents either a drawback of modified gravity theories or a hidden scale-dependence of the parameters.

While the first decade and a half of the 21st century made this issue explicit in the $f(R)$ gravity applications, the last decade has brought the same questioning to the $f(R,T)$ gravity. We have that when working with $f(R)$ gravity, the theory parameters should adapt themselves to the value that the Ricci scalar takes in different systems. For the $f(R,T)$ gravity, if one considers functionals such as $f(R,T)=R+f(T)$, as we did, the theory should adapt its parameters according to the value that $T$ assumes in different systems.

We have investigated the applicability of a particular $f(R,T)$ model in the context of neutron stars. We analyzed the outcomes of the model described by Eq.\eqref{i2}, which was not chosen \textit{ad hoc}, but rather, it was derived using gaussian process regression applied to Hubble data. That is to say that by construction, this functional form is well-succeeded as the basis for a cosmological model, so that our questioning here was if it could also suit the static equilibrium configurations of neutron stars.

We have shown that it is possible to obtain static neutron star configurations from such an $f(R,T)$ model and our main results are presented in Fig.\ref{fig:enter-label}, that shows a slight increase in the predicted maximum mass in comparison with General Relativity results. 

Figure \ref{fig:enter-label2} shows a subtle increase in the energy density of neutron stars in the optimized $f(R,T)$ gravity model when compared to General Relativity Theory outcomes. The behavior of energy density as a function of radius is a crucial aspect for understanding the internal structure of a star. As illustrated in Figure \ref{fig:enter-label2}, the energy density exhibits a maximum value at the core, where gravitational interactions are strongest. This high central energy density profile reflects the balance between gravitational collapse and the pressure exerted by the degenerate neutron gas. In contrast, General Relativity predicts a steeper decline in energy density with increasing radius than the optimized $f(R,T)$ gravity, allowing the latter to display higher densities. This enhanced density capability implies that neutron stars can attain a higher mass that exceeds the limits predicted by General Relativity.  

 Our results indicate that on the compact objects scale, the contributions of the hyperbolic function in $f(T)$, in addition to the linear component on $T$ and the $\gamma$ parameter, are small to guarantee the expected physical behavior of neutron stars. By setting $A = 5.0\times 10^{-4}$ and $\beta = \gamma = 1.05\times 10^{-3}$ we obtain the maximum mass neutron star, with the other parameters in the $f(T)$ functional having the same value as in the cosmological case. 
 
 %less impact, what suggests that these terms have limited influence in stellar-scale systems. On the other hand, it is known that the terms $A$, $\beta T$ and $\gamma$ alone cannot explain the cosmic acceleration in the scenario $f(R,T)$ \cite{velten/2017}. 

%In contrast, the analysis of the $f(T)$ functional suggests that the parameter $\beta$ plays a crucial role at cosmological scales. Fortunato et al. used observational data from the Hubble constant to derive the form of the $f(R,T)$ functional discussed here, obtained $\beta = -2.99$. This suggests that $\beta$ acts as a scale-dependent parameter, being relevant in cosmological regimes and closer to zero at stellar scales.

We conclude that the optimized $f(T)$ functional form, with $A = 5.0\times 10^{-4}$ and $\beta = \gamma = 1.05\times 10^{-3}$ for neutron stars and $A =-1.05\times10^4$, $\beta = -2.99$ and $\gamma =-1.61\times10^4$ for cosmological scales, keeping $\alpha=-1.83\times10^{-5}$,  $\lambda=-2.39\times10^3$ and $T_0=2.58\times10^3$ in both cases, is a robust approach for modeling both local and global phenomena. We encourage further applications in distinct regimes, such as the rotation curves of galaxies, for example, with the purpose of verifying if the optimized $f(R,T)$ gravity is capable of explaining the galactic dynamics with no need for the enigmatic dark matter, and if that is the case, for which values of $A$, $\beta$ and $\gamma$.
 
\begin{acknowledgments}
JTQ and GNZ would like to express their sincere gratitude to Coordenação de Aperfeiçoamento de Pessoal de Nível Superior (CAPES) for the financial support, making this research possible. PHRSM would like to thank CNPq (Conselho Nacional de Desenvolvimento Científico e Tecnológico) for partial financial support under grant No. 310366/2023-2.
\end{acknowledgments}

\end{document}